\documentclass[aps,prd,twocolumn,showpacs,superscriptaddress,nofootinbib,preprintnumbers]{revtex4-1}
\pdfoutput=1
\usepackage{amssymb,amsmath,latexsym,mathrsfs}
\usepackage{graphicx,subfigure}
\usepackage{epsfig}
\usepackage{varioref,xr-hyper}
\usepackage{color}
\usepackage{multirow}
\usepackage{tikz}
\usepackage{array}
\usepackage{hyperref}
\usepackage{wasysym}
\usepackage{color}
\usepackage{float}
\usepackage[utf8]{inputenc}
\usepackage[T1]{fontenc}
\newcommand{\Arrow}[1]{%
\parbox{#1}{\tikz{\draw[->](0,0)--(#1,0);}}
}

\begin{document}

\title{Alive and well: mimetic gravity and a higher-order extension in light of GW170817}

\author{Alessandro Casalino}
\email{alessandro.casalino@unitn.it}
\affiliation{Dipartimento di Fisica, Universit\`{a} di Trento,\\Via Sommarive 14, I-38123 Povo (TN), Italy}
\affiliation{Trento Institute for Fundamental Physics and Applications (TIFPA)-INFN,\\Via Sommarive 14, I-38123 Povo (TN), Italy}

\author{Massimiliano Rinaldi}
\email{massimiliano.rinaldi@unitn.it}
\affiliation{Dipartimento di Fisica, Universit\`{a} di Trento,\\Via Sommarive 14, I-38123 Povo (TN), Italy}
\affiliation{Trento Institute for Fundamental Physics and Applications (TIFPA)-INFN,\\Via Sommarive 14, I-38123 Povo (TN), Italy}

\author{Lorenzo Sebastiani}
\email{lorenzo.sebastiani@unitn.it}
\affiliation{Dipartimento di Fisica, Universit\`{a} di Trento,\\Via Sommarive 14, I-38123 Povo (TN), Italy}
\affiliation{Trento Institute for Fundamental Physics and Applications (TIFPA)-INFN,\\Via Sommarive 14, I-38123 Povo (TN), Italy}

\author{Sunny Vagnozzi}
\email{sunny.vagnozzi@fysik.su.se}
\affiliation{The Oskar Klein Centre for Cosmoparticle Physics, Stockholm University, Roslagstullsbacken 21A, SE-106 91 Stockholm, Sweden}
\affiliation{The Nordic Institute for Theoretical Physics (NORDITA), Roslagstullsbacken 23, SE-106 91 Stockholm, Sweden}

\date{\today}

\begin{abstract}
The near-simultaneous multi-messenger detection of the gravitational wave (GW) event GW170817 and its optical counterpart, the short $\gamma$-ray burst GRB170817A, implies that deviations of the GW speed from the speed of light are restricted to being of ${\cal O}(10^{-15})$. In this note, we study the implications of this bound for mimetic gravity and confirm that in the original setting of the theory, GWs propagate at the speed of light, hence ensuring agreement with the recent multi-messenger detection. A higher-order extension of the original mimetic theory, appearing in the low-energy limit of projectable Ho\v{r}ava-Lifshitz gravity, is then considered. Performing a Bayesian statistical analysis where we compare the predictions of the higher-order mimetic model for the speed of GWs against the observational bound from GW170817/GRB170817A, we derive constraints on the three free parameters of the theory. Imposing the absence of both ghost instabilities and superluminal propagation of scalar and tensor perturbations, we find very stringent 95\% confidence level upper limits of $\sim 7 \times 10^{-15}$ and $\sim 4 \times 10^{-15}$ on the coupling strengths of Lagrangian terms of the form $\nabla^{\mu}\nabla^{\nu}\phi\nabla_{\mu}\nabla_{\nu}\phi$ and $(\Box \phi)^2$ respectively, with $\phi$ the mimetic field. We discuss implications of the obtained bounds for mimetic theories. This work presents the first ever robust comparison of a mimetic theory to observational data.
\end{abstract}

\pacs{}
\maketitle

\section{Introduction}
The recent joint detection of the gravitational wave (GW) event GW170817 by the LIGO/Virgo collaboration~\cite{TheLIGOScientific:2017qsa}, and of its optical counterpart, the short $\gamma$-ray burst GRB170817A, by the \textit{Fermi} Gamma-ray Burst Monitor and the Anti-Coincidence Shield for the \textit{INTEGRAL} spectrometer~\cite{Monitor:2017mdv}, has inaugurated the era of multi-messenger astronomy. The near-simultaneous detection of the two signals implies that GWs travel at a speed $c_T$ which is nearly the speed of light~\cite{Monitor:2017mdv}. Many exotic theories of gravity feature an effective cosmological medium which spontaneously breaks Lorentz invariance, implying that GWs (the excitations of the medium) no longer travel at the speed of light~\cite{Bettoni:2016mij} (see also~\cite{Lombriser:2015sxa,deRham:2016wji}). As a consequence, several previously theoretically motivated modified gravity theories~\cite{Nojiri:2006ri,Nojiri:2010wj,Clifton:2011jh,Capozziello:2011et,Nojiri:2017ncd} are no longer viable~\cite{Creminelli:2017sry,Sakstein:2017xjx,Ezquiaga:2017ekz,Baker:2017hug,Ezquiaga:2018btd} (see~\cite{Lombriser:2015sxa,Lombriser:2016yzn} for important early work, and~\cite{Boran:2017rdn,Nojiri:2017hai,Arai:2017hxj,Amendola:2017orw,Visinelli:2017bny,Crisostomi:2017lbg,
Langlois:2017dyl,Gumrukcuoglu:2017ijh,Kreisch:2017uet,Bartolo:2017ibw,Dima:2017pwp,Cai:2018rzd,Pardo:2018ipy}).

A particularly interesting modified gravity theory is represented by mimetic gravity (MimG). Proposed in 2013 by Chamseddine and Mukhanov~\cite{Chamseddine:2013kea}~\footnote{See however~\cite{Lim:2010yk,Capozziello:2010uv,Gao:2010gj,Zumalacarregui:2013pma} for important earlier work.}, the theory is related to General Relativity via a non-invertible disformal transformation of the metric, involving a mimetic scalar field $\phi$~\cite{Deruelle:2014zza,Domenech:2015tca,Achour:2016rkg}. The non-trivial vacuum solutions of the theory effectively mimic cold dark matter (DM) on cosmological scales, while simple generalizations of the original model can mimic any given cosmological evolution~\cite{Chamseddine:2014vna} (see~\cite{Sebastiani:2016ras} for a recent review).

As recently noticed in~\cite{Ezquiaga:2017ekz}, MimG is a particular case of a degenerate higher order scalar-tensor (DHOST) theory~\cite{Zumalacarregui:2013pma,Langlois:2015cwa}~\footnote{See also~\cite{Achour:2016rkg,Langlois:2018jdg} for related discussions on the relation between MimG and DHOST theories.} obtained by conformally transforming a Horndeski theory with $c_T=1$~\footnote{We use natural units, where the speed of light takes the value $1$.}: this implies that $c_T=1$ also in MimG, ensuring the consistency of the original theory with GW170817. It is unclear, however, whether this conclusion carries over to the many proposed extensions of MimG. Given the interest spurred by MimG in the community, addressing this issue is important and timely.

In this note, it is our aim to study the viability of extended mimetic gravity models in light of GW170817. We focus on a particular model proposed by Cognola \textit{et al.} in~\cite{Cognola:2016gjy}. Our choice is motivated by theoretically appealing properties of the model, which can be viewed as the low-energy limit of projectable Ho\v{r}ava-Lifshitz gravity. We show that GW170817 sets stringent constraints on a term of the form $\nabla^{\mu}\nabla^{\nu}\phi\nabla_{\mu}\nabla_{\nu}\phi$ in the action. Performing a Bayesian statistical analysis where the predictions of the model are compared against observational data from GW170817, we obtain constraints on the three free parameters of the model. This represents the \textit{first time} that a mimetic theory is robustly constrained against observational data.

This note is then organized as follows. In Sec.~\ref{sec:mim}, we very briefly review mimetic gravity and its extensions, and in Sec.~\ref{sec:per} we define the mimetic model we will be considering and study scalar and tensor perturbations within this model, as well as introduce the constraints on tensor perturbations imposed by the GW170817/GRB170817A detection. In Sec.~\ref{sec:method}, we describe our analysis methodology, before proceeding to Sec.~\ref{sec:results} where we discuss our results. Finally, we conclude in Sec.~\ref{sec:conclusions}.

\section{Mimetic gravity and its extensions}
\label{sec:mim}
Mimetic gravity can be obtained by starting from the Einstein-Hilbert (EH) action and reparametrizing the physical metric $g_{\mu \nu}$ in terms of an auxiliary metric $\tilde{g}_{\mu \nu}$ and the mimetic field $\phi$:
\begin{eqnarray}
g_{\mu \nu} = -\tilde{g}_{\mu \nu}\tilde{g}^{\alpha\beta}\partial_{\alpha}\phi\partial_{\beta}\phi \, .
\label{mimetictransformation}
\end{eqnarray}
When varying the action with respect to $g_{\mu \nu}$, taking into account its dependence on $\tilde{g}_{\mu \nu}$ and $\phi$ through Eq.~(\ref{mimetictransformation}), the resulting gravitational field equations feature an extra term which, on a flat FLRW background, behaves as a pressureless fluid and hence mimics cold DM~\cite{Chamseddine:2013kea}. The gradient of the mimetic field is required, for consistency, to satisfy the condition $X \equiv g^{\mu\nu}\partial_{\mu}\phi\partial_{\nu}\phi/2=-1/2$, which can be implemented as a constraint at the level of the action through a Lagrange multiplier term $\lambda$~\cite{Chamseddine:2014vna}. Various works focusing on cosmological and astrophysical issues within MimG have been conducted in the subsequent literature, for an incomplete list see e.g.~\cite{Golovnev:2013jxa,Barvinsky:2013mea,Nojiri:2014zqa,Saadi:2014jfa,Capela:2014xta,Mirzagholi:2014ifa,Leon:2014yua,
Haghani:2015iva,Matsumoto:2015wja,Momeni:2015gka,Myrzakulov:2015sea,Astashenok:2015haa,Myrzakulov:2015qaa,
Myrzakulov:2015kda,Odintsov:2015wwp,Hammer:2015pcx,Ramazanov:2016xhp,Nojiri:2016ppu,Nojiri:2016vhu,Sadeghnezhad:2017hmr,
Baffou:2017pao,Vagnozzi:2017ilo,Bouhmadi-Lopez:2017lbx,Shen:2017rya,Nojiri:2017ygt,Dutta:2017fjw,
Brahma:2018dwx,deHaro:2018sqw,Zhong:2018tqn,Chamseddine:2018qym,Chamseddine:2018gqh,Firouzjahi:2018xob,Zlosnik:2018qvg,
Gorji:2018okn,Nashed:2018qag,Ganz:2018vzg}.

The original MimG theory was constructed starting from a ``seed'' EH action, and various extensions of the theory have been considered by starting from different ``seed'' actions. Mimetic Horndeski gravity uses Horndeski's theory, the most general 4-dimensional scalar-tensor theory of gravity with second-order field equations~\cite{Horndeski:1974wa}, as ``seed'' theory~\cite{Arroja:2015wpa}, and has received particular interest recently~\cite{Rabochaya:2015haa,Arroja:2015yvd,Cognola:2016gjy,Arroja:2017msd}.

In the original MimG model, the sound speed $c_s$ (i.e. the speed of propagation of scalar perturbations) is identically $0$~\cite{Chamseddine:2014vna}. This is problematic if one wants to define quantum fluctuations in the mimetic field. Later it was shown that the problem actually persists also in mimetic Horndeski gravity~\cite{Arroja:2015yvd}. In~\cite{Cognola:2016gjy}, some of us studied an explicit mimetic Horndeski model, from which we then constructed a higher-order mimetic model by explicitly breaking the original Horndeski structure in order to achieve $c_s \neq 0$. In this note, we will be considering this higher-order mimetic model, which for simplicity we refer to as HOMim model.

\section{Scalar and tensor perturbations in the HOMim model}
\label{sec:per}
The action of the HOMim model is given by~\cite{Cognola:2016gjy}~\footnote{Note that the same model was later studied in~\cite{jirousekthesis}, see also~\cite{Rinaldi:2016oqp,Diez-Tejedor:2018fue} for related studies.}:
\begin{eqnarray}
S = \frac{1}{2}\int d ^4x\sqrt{-g} [ R(1 + a g^{\mu \nu}\nabla_{\mu}\phi\nabla_{\nu}\phi) - \frac{c}{2}(\Box \phi)^2 \nonumber \\
+ \frac{b}{2} (\nabla_{\mu}\nabla_{\nu}\phi)^2 - \frac{\lambda}{2} \left ( g ^{\mu \nu}\nabla _{\mu}\phi\nabla _{\nu}\phi + 1 \right )  - V (\phi) ] \,.
\label{cmsvz}
\end{eqnarray}
Setting $b=c=4a$ in the action recovers the Horndeski structure of the theory. The equations of motion of the theory can be found in our companion paper~\cite{Casalino:2018tcd}. In~\cite{Cognola:2016gjy}, it was argued that the model can be viewed as the low-energy limit of projectable Ho\v{r}ava-Lifshitz gravity~\cite{Horava:2009uw}, a power-counting renormalizable candidate theory of quantum gravity, hence lending to its theoretical appeal~\footnote{The mimetic Horndeski model from which the HOMim action was constructed was inspired by previous work in the context of ``covariant renormalizable gravity'' models~\cite{Nojiri:2009th,Nojiri:2010tv,Cognola:2010by,Nojiri:2010kx,Chaichian:2011sx}, wherein power-counting renormalizability is achieved by breaking Lorentz invariance dynamically. In this way, theoretical issues present in Ho\v{r}ava-Lifshitz gravity and connected to the explicit breaking of diffeomorphism invariance~\cite{Blas:2009yd,Blas:2009qj,Blas:2009ck} are avoided.}.

By perturbing a background FLRW line element appropriately, we have derived the \textit{sound speed} $c_s$ and the \textit{gravitational wave speed} $c_T$. Here we simply quote the result and refer the reader to~\cite{Casalino:2018tcd} for further details on the calculation~\footnote{Notice that $c_s$ and $c_T$ do not depend on the scalar field derivative $\dot{\phi}$, unlike what happens in Horndeski gravity, because in mimetic gravity the Lagrange multiplier constraint fixes $\dot{\phi}=1$ on a FLRW background.}:
\begin{eqnarray}
\label{cs}
c_s^2&=&\frac{2(b-c)(a-1)}{(2a-b-2)(4-4a-b+3c)}\,, \\
c_{T}^{2}&=&\frac{2(1-a)}{2(1-a)+b}\,.
\label{ct}
\end{eqnarray}
From Eq.~(\ref{ct}) it is clear that $b\neq0$ is a necessary condition for obtaining $c_T\neq1$. We see that bounds on the GW speed from GW170817 will constrain the parameters $a$ and $b$, whereas further information on the sound speed is necessary in order to put constraints on $c$. Notice also that, when $a=b=c=0$, we recover $c_s^2=0$ and $c_T^2=1$, in agreement with expectations from the original MimG model, and in full agreement with the GW170817/GRB170817A detection.

The equations of motion derived form the action in Eq.~(\ref{cmsvz}) are of at most fourth order, as shown in the companion paper~\cite{Casalino:2018tcd}. In particular the Klein-Gordon equation, Eq.~(5) in~\cite{Casalino:2018tcd}, is of fourth order. However, by imposing the constraint coming from the Lagrange multiplier $\lambda$ (Eq.~(6) in~\cite{Casalino:2018tcd}) and by choosing a FLRW metric, we obtain equations of motion of second order in the Hubble rate $H$, see Eqs.~(9-11) in~\cite{Casalino:2018tcd}. Hence, at the background level there do not appear to be ghosts. At the perturbative level, the situation is more delicate, since as soon as the sound speed is non-zero there is an extra propagating scalar mode, for a total of three degrees of freedom. We have discussed this issue in more detail in~\cite{Casalino:2018tcd}, where we have computed the quadratic action of the scalar and tensor perturbations, and found that the theory is free of ghosts only when choosing $a<1$ and $c>0$. Heretofore, we shall impose these conditions in order to ensure the theoretical consistency of the model. In addition to these conditions ensuring the absence of ghosts, stability arguments impose the conditions $0 \leq c_s^2 \leq 1$ and $0 \leq c_T^2 \leq 1$. The upper limit of $1$ on $c_s^2$ and $c_T^2$ enforces the absence of superluminal propagation of scalar and tensor modes. It has been shown that superluminality in a theory of gravity is incompatible with an UV-complete theory whose S-matrix satisfies canonical analyticity constraints~\cite{Adams:2006sv,Salvatelli:2016mgy}.

The recent near-simultaneous detection of GW170817~\cite{TheLIGOScientific:2017qsa} and its optical counterpart GRB170817A~\cite{Monitor:2017mdv}, has placed very stringent constraints on $\delta c_T$, the fractional deviation of the GW speed from the speed of light. Following~\cite{Ezquiaga:2017ekz}, we will consider the following bound:
\begin{eqnarray}
\left \vert \delta c_T \right \vert < 5\times 10^{-16} \, .
\label{boundsymmetric}
\end{eqnarray}

The bound in Eq.~(\ref{boundsymmetric}) provides a very stringent constraint on deviations of $c_T$ from the speed of light (note that the region $c_T<1$ was already previously constrained by non-observation of gravi-\v{C}erenkov radiation from cosmic rays, although the bound is not competitive with the GW170817/GRB170817A one~\cite{Moore:2001bv}). Let us imagine for a moment \textit{fixing} the requirement $c_T=1$, which is trivially satisfied in the baseline MimG model [setting $a=b=0$ in Eq.~(\ref{ct})]. In the context of the HOMim mimetic model, from Eq.~(\ref{ct}) it follows that $c_T \equiv 1$ instead imposes the \textit{very stringent} constraint $b=0$. This implies that a term of the form $\nabla^{\mu}\nabla^{\nu}\phi\nabla_{\mu}\nabla_{\nu}\phi$ is prohibited from appearing in the HOMim action. In the remaining part of the work, we will entertain the possibility of a tiny violation of the constraint $c_T \equiv 1$, in accordance with the the bounds on $c_T$ provided by Eq.~(\ref{boundsymmetric}), and explore the implications of this bound on the parameters of the HOMim model.

\section{Analysis methodology}
\label{sec:method}
We perform a standard Bayesian statistical analysis to constrain the three parameters of the extended mimetic model ($a$, $b$, and $c$) in light of the near-simultaneous GW170817/GRB170817A detection. The constraint on $\delta c_T$ of Eq.~(\ref{boundsymmetric}) can only be used to provide bounds on $a$ and $b$ [Eq.~(\ref{ct})]. The first part of our analysis is therefore concerned with determining the joint and marginalized posterior probability distributions of $a$ and $b$, in light of observational data $\boldsymbol{d}$ given the constraint on $\delta c_T$ of Eq.~(\ref{boundsymmetric}).

We begin by considering the parameters $\boldsymbol{\theta} \equiv (a,\,b)$. To proceed, we need to construct the likelihood ${\cal L}(\boldsymbol{\theta})$, consisting of the probability of observing the data $\boldsymbol{d}$ given a choice of model parameters $\boldsymbol{\theta}$: ${\cal L}(\boldsymbol{\theta}) = P(\boldsymbol{d} \vert \boldsymbol{\theta})$. Following Eq.~(\ref{boundsymmetric}), we model the likelihood as an univariate Gaussian in $\delta c_T$, centered around $\delta c_T=0$:
\begin{eqnarray}
{\cal L}(\boldsymbol{\theta}) \equiv {\cal L}(a,b) = \exp \left \{ -\frac{ \left [ \delta c_T(a,b) \right ]^2}{2\sigma_{\delta c_T}^2} \right \} \, ,
\label{likelihood}
\end{eqnarray}
where $\delta c_T(a,b)$, following Eq.~(\ref{ct}), is given by:
\begin{eqnarray}
\delta c_T(a,b) = \sqrt{\frac{2(1-a)}{2-2a+b}}-1\,.
\end{eqnarray}
Finally, in Eq.~(\ref{likelihood}), $\sigma_{\delta c_T}$ denotes the uncertainty on $\delta c_T$, which we estimate as $\sigma_{\delta c_T} = 5\times10^{-16}$ following Eq.~(\ref{boundsymmetric}).

Using Bayes' theorem, we construct the joint posterior distribution of $a$ and $b$ as the product of the likelihood [Eq.~(\ref{likelihood})] and the prior probability distributions we assign to $a$ and $b$. The choice of prior is dictated by a combination of theoretical and phenomenological considerations. Following our previous discussion, we first of all impose the requirement of subluminality of tensor perturbations: $c_T^2(a,b) \leq 1$.

In the action Eq.~(\ref{cmsvz}), the term multiplying the Riemann tensor is $1+ag^{\mu \nu}\nabla_{\mu}\phi\nabla_{\nu}\phi = 1+2aX = 1-a$. As this term controls the strength of the effective Newton constant, we must impose its non-negativity, which implies $a<1$. Notice that, as discussed in Sec.~\ref{sec:per}, requiring the absence of ghosts anyway led to the condition $a<1$. In addition, guided by perturbativity arguments, we expect $\vert a \vert \lesssim {\cal O}(1)$, as in general it could be problematic to embed a coupling constant $\vert a \vert \gg{\cal O}(1)$ in the context of a UV-complete theory of gravity. Guided by these considerations, we choose for simplicity to impose a top-hat (flat) prior on $a$ within the range $[-1,1]$. We assess a posteriori that our results are only mildly affected by other choices of flat prior as long as the upper and lower limits are $\sim {\cal O}(1)$ in modulo.

Concerning $b$, we already know that this parameter is required to be $\ll {\cal O}(1)$, in order for the bound in Eq.~(\ref{boundsymmetric}) to be satisfied. Moreover, we see from Eq.~(\ref{ct}) that for $b<0$, one would obtain $c_T^2>1$, which violates the subluminality requirement. Based on these arguments, we impose a top-hat prior on $b$ within the range $[0,1]$. In conclusion, the joint posterior distribution of $a$ and $b$ we sample from is given by:
\begin{eqnarray}
&&P(a,b \vert \boldsymbol{d}) = \exp \left \{ -\frac{ \left [ \delta c_T(a,b) \right ]^2}{2\sigma_{\delta c_T}^2} \right \} \Theta(c_T^2)\Theta(1-c_T^2) \nonumber \\
&&\times \Theta(1+a)\Theta(1-a)\Theta(b)\Theta(1-b) \, ,
\label{posteriorab}
\end{eqnarray}
where $\Theta(x)$ denotes the Heaviside step function.

In the second part of the analysis we include the parameter $c$ as well, which requires additional information to be taken into account beyond the constraint on $\delta c_T$ of Eq.~(\ref{boundsymmetric}). Since $c$ enters in the expression for the sound speed $c_s$ [Eq.~(\ref{cs})], we additionally impose the subluminality of scalar perturbations~\footnote{There exist upper limits on the sound speed of DM from observations of the CMB and large-scale structure, which suggest $c_s^2 \lesssim 10^{-10.7}$~\cite{Kunz:2016yqy}. However, these bounds are not entirely model-independent: the analysis should be re-performed for the HOMim model, by solving the relevant Einstein-Boltzmann equations. Moreover, the propagating scalar mode in the HOMim model does not exclusively mimic DM, but a combination of DM and dark energy (see~\cite{Casalino:2018tcd}). In order to be conservative, we have decided to not impose these upper limits on $c_s$.}. In addition, as discussed in Sec.~\ref{sec:per}, requiring the absence of ghosts leads to the condition $c>0$. Therefore, guided by considerations on the absence of ghosts and perturbativity as per our previous discussion, we impose a top-hat prior on $c$ within the range $[0,1]$. We will later anyway see that data require $c \ll {\cal O}(1)$. In this case, the joint posterior distribution of $a$, $b$, and $c$, given the data $\boldsymbol{d}$, is given by:
\begin{eqnarray}
&&P(a,b,c \vert \boldsymbol{d}) = \exp \left \{ -\frac{ \left [ \delta c_T(a,b) \right ]^2}{2\sigma_{\delta c_T}^2} \right \} \nonumber \\
&&\times \Theta(1+a)\Theta(1-a)\Theta(b)\Theta(1-b)\Theta(c)\Theta(1-c) \nonumber \\
&&\times \Theta(c_T^2)\Theta(1-c_T^2)\Theta(c_s^2)\Theta(1-c_s^2) \, .
\label{posteriorabc}
\end{eqnarray}

To sample the posterior distributions of Eq.~(\ref{posteriorab}) and Eq.~(\ref{posteriorabc}), we make use of Markov Chain Monte Carlo (MCMC) methods, by implementing the Metropolis-Hastings algorithm. We use the cosmological MCMC sampler \texttt{Montepython}~\cite{Audren:2012wb}, configured to act as a generic sampler. We monitor the convergence of the MCMC chains using the Gelman and Rubin parameter $R-1$~\cite{Gelman:1992zz}, which we require to be $<0.01$ in order for the chains to be considered converged.

\section{Results}
\label{sec:results}
\begin{figure}[!t]
\includegraphics[width=1.0\linewidth]{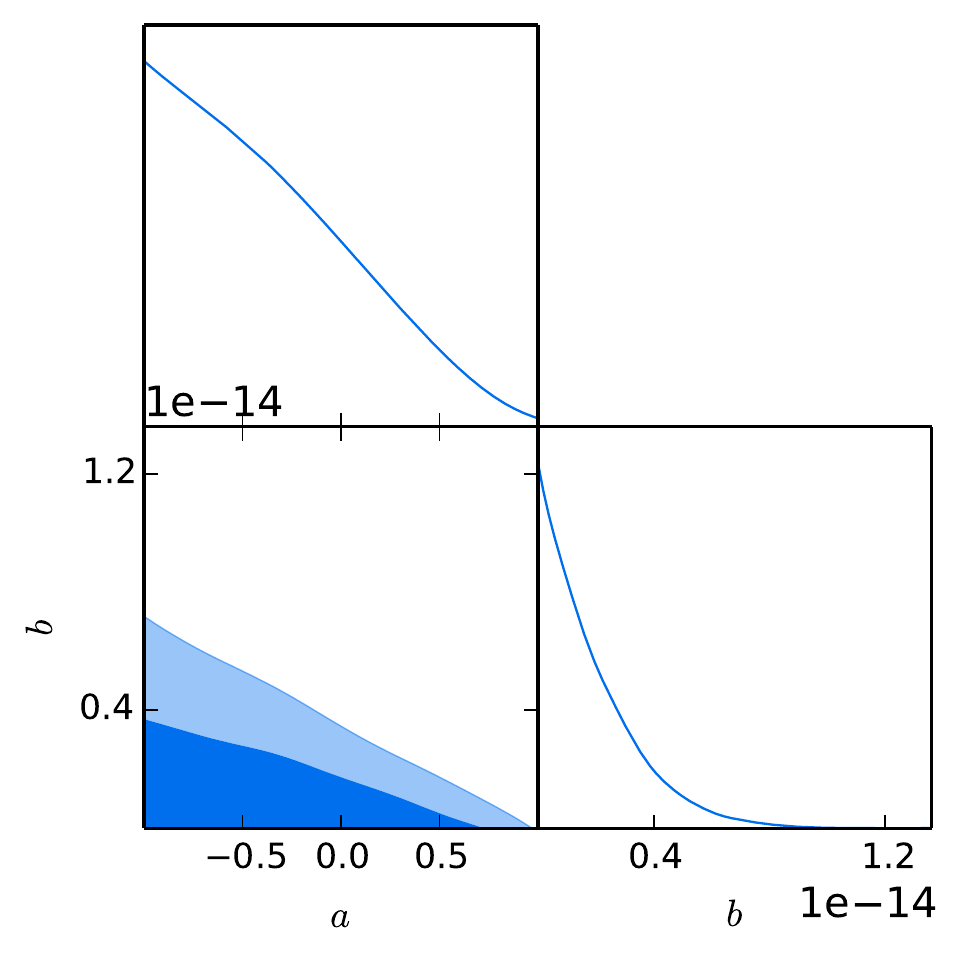}
\caption{Triangular plot showing joint and marginalized posterior probability distributions of the parameters $a$ and $b$, in light of the joint GW170817/GRB170817A detection, and imposing the subluminality of tensor perturbations and absence of ghosts. The blocks along the diagonal (upper left and lower right) contain the 1D marginalized posterior distributions of $a$ and $b$ (since the quantity plotted is a \textit{normalizable} probability distribution, the overall scale of these plots is irrelevant, which is the reason why the $y$-axes are unit-less). The remaining block (lower left) shows the joint 2D $a$-$b$ posterior distribution, with 68\%~C.L. and 95\%~C.L. credible regions corresponding to the light and dark blue areas respectively.}
\label{fig:ab}
\end{figure}

We first sample the joint $a$-$b$ posterior distribution given by Eq.~(\ref{posteriorab}). We show the results in the triangular plot of Fig.~\ref{fig:ab}, whose diagonal contains the marginalized probability distributions of the two parameters.

We find $a<0.55$ at 95\% confidence level (C.L.), while the marginalized posterior of $b$ is, as expected, peaked at $b=0$ and falls rapidly as $b$ increases, indicating $b<5.11 \times 10^{-15}$ at 95\%~C.L.. The reason for these very tight bounds is readily found by inspecting Eq.~(\ref{ct}). As $b$ moves away from $0$ (at fixed $a$), $c_T^2$ rapidly moves away from $1$, and hence the probability density of the given point in $(a,b)$ parameter space decreases.

Although deviations of $c_T$ from $1$ are mostly controlled by $b$, the parameter $a$ does nonetheless play a role. In fact, from the orientation of the joint $a$-$b$ posterior distribution (lower left panel in Fig.~\ref{fig:ab}), we see that the two parameters exhibit a mild negative correlation (also referred to as parameter degeneracy). That is, it is possible to increase/decrease one parameter and correspondingly decrease/increase the other, and still maintain consistency with the GW170817 bound on $c_T$. This observation can be rigorously shown by Taylor expanding $\delta c_T$ in the limit of small $b/(2-2a)$:
\begin{eqnarray}
\delta c_T = \left ( 1+\frac{b}{2-2a} \right )^{-\frac{1}{2}}-1 \underset{\frac{b}{2-2a} \to 0}{\Arrow{1cm}} -\frac{b}{4(1-a)} \, .
\label{taylor}
\end{eqnarray}
From Eq.~(\ref{taylor}), we see that the more $b$ increases, the more $c_T$ deviates from $1$. Moreover, the smaller $a$ is, the larger the term $4(1-a)$ in Eq.~(\ref{taylor}) is, implying that it is consequently possible to ``tolerate'' larger values of $b$ and still be consistent with the deviation of $c_T$ from $1$ allowed by GW170817/GRB170817A. This explains the mild negative correlation between $a$ and $b$. From our MCMC chains we estimate the correlation coefficient between the two parameters to be $\approx -0.40$.

\begin{figure}[!t]
\includegraphics[width=1.0\linewidth]{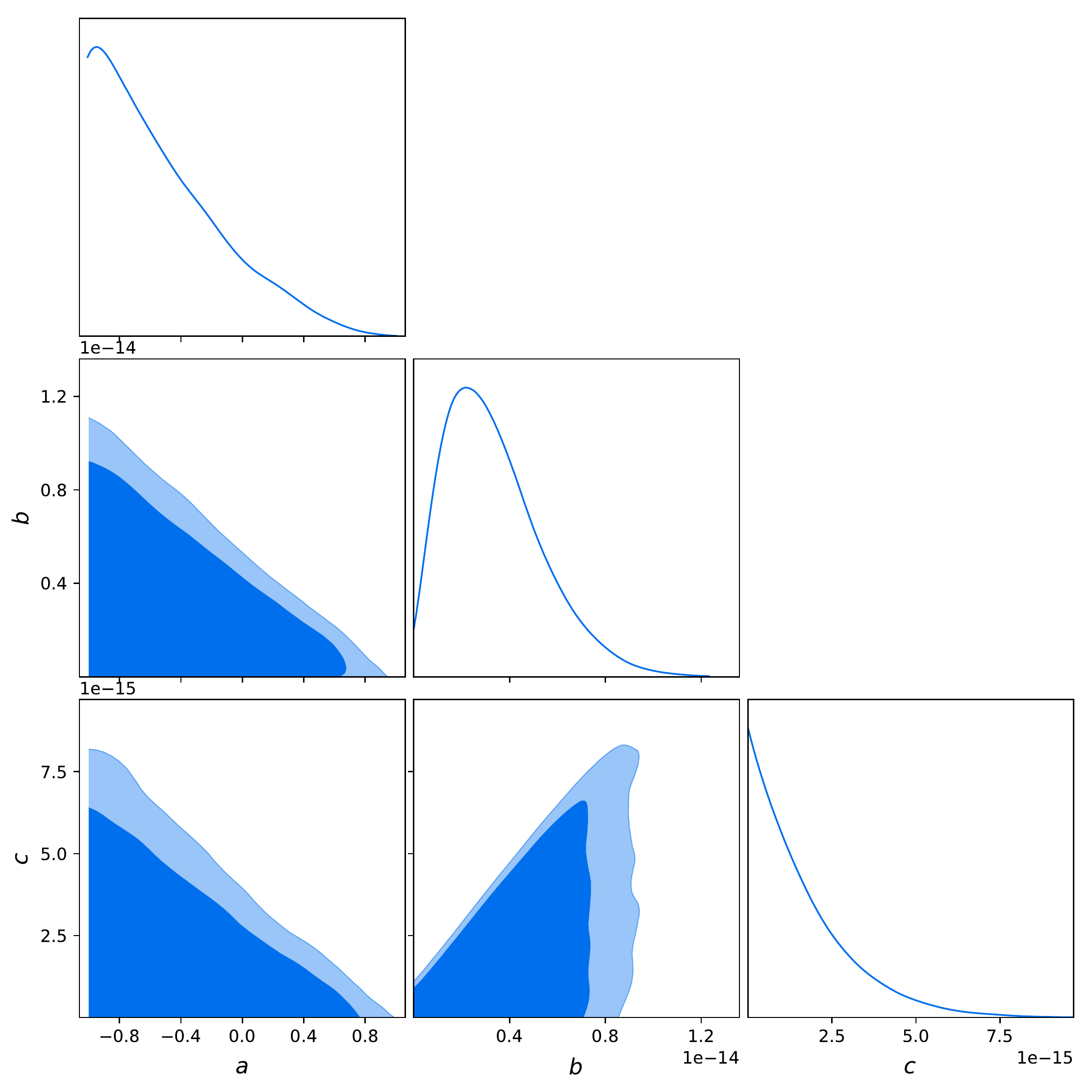}
\caption{As in Fig.~\ref{fig:ab}, with the addition of the parameter $c$ and the further imposition of subluminality of scalar perturbations.}
\label{fig:abc}
\end{figure}

Next, we sample the joint $a$-$b$-$c$ posterior probability distribution given by Eq.~(\ref{posteriorabc}). We show the results in the triangular plot of Fig.~\ref{fig:abc}. Quoting all 95\%~C.L. upper bounds, we find $a<0.27$, $b<7.18 \times 10^{-15}$, and $c<4.68 \times 10^{-15}$.~\footnote{Although there appears to be a mild peak in the posterior distribution of $b$, we find that the distribution is consistent with $b=0$ at $\sim 2 \sigma$. Therefore, as is standard practice in the field, we only quote an upper limit for $b$ instead of a ``detection" of non-zero $b$. Recall also that we had chosen the upper and lower limits for our priors based on perturbativity considerations. We have checked that our results, and in particular the upper limits on $b$ on $c$, are only very marginally affected (within the same order of magnitude) by other choices for the upper and lower limits of the prior which still are ${\cal O}(1)$ in modulo. We therefore consider our results relatively robust against the choice of prior.} In order to explain the results we find, it is useful to consider the expression for the sound speed squared, Eq.~(\ref{cs}), and the combinations of the three parameters $a$, $b$, and $c$ necessary to keep this quantity positive. It is then quite easy to see that, in the limit where $b \ll a$ and $c>0$, it is possible to keep $c_s^2 \geq 0$ by requiring that $c$ be smaller than $b$, i.e. $c \lesssim b \sim O(10^{-15})$, while also having $1-a \gg O(10^{-15})$ (i.e. $a$ is sufficiently far from $1$). In this limit, the sound speed is approximately given by:
\begin{eqnarray}
c_s^2(a,b,c) \approx \frac{(b-c)}{4-4a-b+3c} \,,
\label{approximating}
\end{eqnarray}
where the condition $c \lesssim b \sim O(10^{-15})$ now ensures that the numerator  of Eq.~(\ref{approximating}) is positive, while the condition $(1-a) \gg O(10^{-15})$ ensures that there are no ``accidental cancellations"  between the terms $4(1-a)$ and $3c-b$ in the denominator which might otherwise make it negative, i.e. that the denominator is approximately given by $4(a-1)$ and hence is always positive since $a<1$. This discussion explains why the upper limit on $c$ is approximately of the same order of the upper limit on $b$, i.e. of order $10^{-15}$, since $c$ is required to be positive (to avoid ghosts) but smaller than $b$ (to have $c_s^2 \geq 0$).

\begin{figure}[!t]
\includegraphics[width=1.0\linewidth]{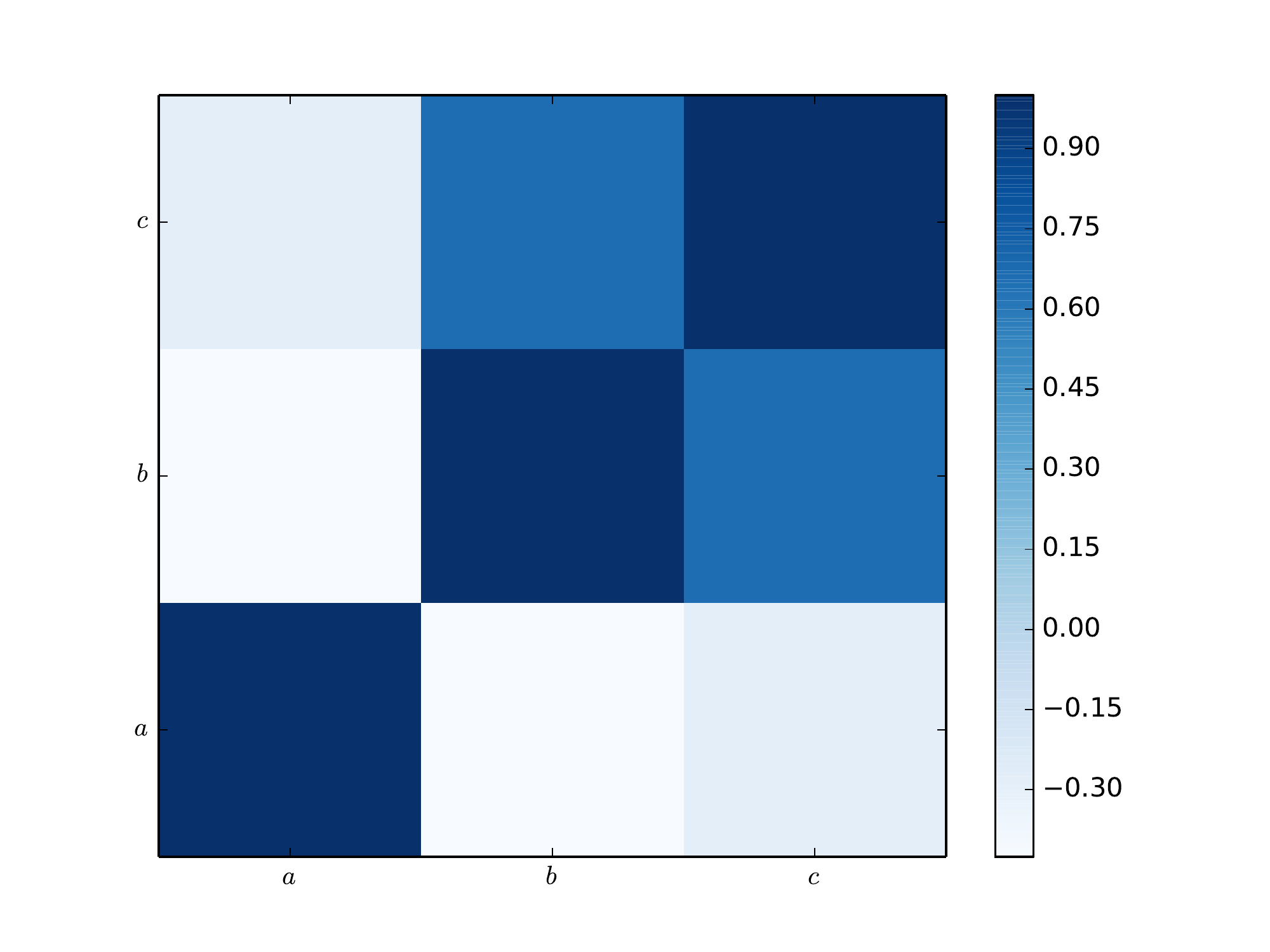}
\caption{Heatmap of the correlation matrix between the 3 parameters ($a$, $b$, and $c$) we are examining. We visually see that the strongest correlation is that between $b$ and $c$, resulting from the necessity of avoiding ghost instabilities (which requires $c>0$) while needing $c_s^2 \geq 0$ (which requires $c<b$), as discussed in the text.}
\label{fig:correlation}
\end{figure}

The above discussion also suggests that we can expect a strong positive correlation between $b$ and $c$ (since increasing $c$ requires increasing $b$ in order to keep the numerator of Eq.~(\ref{approximating}) positive). In fact, we find a correlation coefficient of $0.66$ between $b$ and $c$, which is stronger than the already strong correlation we previously found between $a$ and $b$. On the other hand, we find a weaker correlation between $a$ and $c$, with correlation coefficient of $-0.28$, which is induced by the mutual correlations of these two parameters with $b$. The negative correlation between $a$ and $c$, and the positive one between $b$ and $c$, explain why introducing the parameter $c$ has respectively tightened and loosened the upper limits we previously derived on $a$ and $b$ when only considering these two parameters (recall that the upper limit shifted from $0.55$ to $0.27$ for $a$, and from $5.11 \times 10^{-15}$ to $7.18 \times 10^{-15}$ for $b$). We plot a heatmap of the correlation coefficients between the 3 parameters in Fig.~\ref{fig:correlation}, where it is clear that the strongest correlation is that between $b$ and $c$, for reasons already discussed previously.

\section{Conclusions}
\label{sec:conclusions}
In this note, we have examined the status of mimetic gravity in light of the recent near-simultaneous detection of the GW event GW170817~\cite{TheLIGOScientific:2017qsa} and its optical counterpart, the short $\gamma$-ray burst GRB170817A~\cite{Monitor:2017mdv}. The original theory~\cite{Chamseddine:2013kea} is in perfect agreement with this multi-messenger detection, as the speed of GWs $c_T$ is therein identically equal to the speed of light: $c_T=1$.

We have then considered a theoretically motivated extended higher-order mimetic model [the HOMim model, Eq.~(\ref{cmsvz})], which appears in the low-energy limit of projectable Ho\v{r}ava-Lifshitz gravity~\cite{Cognola:2016gjy}. Entertaining the possibility of a tiny violation of the $c_T \equiv 1$ constraint, in agreement with experimental constraints from GW170817/GRB170817A [Eq.~(\ref{ct})]~\cite{Monitor:2017mdv}, we have  performed a Bayesian statistical analysis to derive observational constraints on the three free parameters of the HOMim model. In particular, we have found that $b$ and $c$, the coefficients of the terms $\nabla^{\mu}\nabla^{\nu}\phi\nabla_{\mu}\nabla_{\nu}\phi$ and $(\Box \phi)^2$ in the action respectively (with $\phi$ the mimetic field), are subject to the very stringent constraints $0 \leq b < 7.18 \times 10^{-15}$ and $0 \leq c < 4.68 \times 10^{-15}$ at 95\% confidence level. In light of these very tight limits it is tempting to conclude that, to avoid incurring in fine-tuning and naturalness issues, both parameters should in fact be $0$.~\footnote{See instead~\cite{Casalino:2018tcd} for a brief discussion concerning another possibility, namely that the parameters $b$ and $c$ might still be small but non-zero, and might arise as quantum corrections~\cite{Brouzakis:2013lla,Saltas:2016nkg,Saltas:2016awg}.} In this case, the gravitational wave speed is identically equivalent to the speed of light, while the sound speed squared is $1/4(1-a)$ and always positive as long as $a<1$, which is required to avoid ghost instabilities. We stress that in our paper it is the \textit{first time} that robust observational constraints are placed on a mimetic theory.

We conclude mentioning theoretical issues pertaining the instability of the theory. It has recently been argued that mimetic gravity and its higher-order extensions suffer from gradient and ghost instability issues, which undermine the theoretical consistency of the theory~\cite{Chaichian:2014qba,Ijjas:2016pad,Firouzjahi:2017txv,Yoshida:2017swb,Hirano:2017zox,Zheng:2017qfs,Cai:2017dyi,
Cai:2017dxl,Takahashi:2017pje,Gorji:2017cai}~\footnote{Although in principle imposing $c_s^2\geq0$ and $c_T^2\geq0$ as we did should at least alleviate the gradient instability issues.}. Although in this work we have imposed specific conditions on the parameters to avoid ghost instabilities, we notice that more generally there exist paths to curing these instabilities, for instance by considering direct couplings between higher derivatives of the mimetic field and curvature~\cite{Hirano:2017zox,Zheng:2017qfs,Cai:2017dyi,Takahashi:2017pje,Gorji:2017cai}. One would in general expect such couplings to lead to deviations of $c_T$ from the speed of light, rendering the theory at odds with observational constraints, and warranting a new analysis along the lines of the one conducted here. Other future research directions could include constraining mimetic gravity from the study of primordial gravitational waves, \textit{i.e.} from measurements of the cosmic microwave background BB spectrum, along the lines of the study performed in~\cite{Nunes:2018evm} for the case of $f(T)$ gravity. We defer these studies to future work. \\

\begin{acknowledgments}
We thank Guido Cognola and Sergio Zerbini for useful discussions and collaboration in the initial stages of the project. We thank the referee for very useful comments which helped us improve the quality of the paper. S.V. thanks Pavel Jirou\v{s}ek for useful correspondence, and Thejs Brinckmann for help with the \texttt{Montepython} software.
\end{acknowledgments}

\bibliographystyle{JHEP}
\bibliography{gw.bib}

\providecommand{\href}[2]{#2}\begingroup\raggedright\begin{thebibliography}{100}

\bibitem{TheLIGOScientific:2017qsa}
{\scshape Virgo, LIGO Scientific} collaboration, B.~Abbott et~al.,
  \emph{{GW170817: Observation of Gravitational Waves from a Binary Neutron
  Star Inspiral}},
  \href{https://doi.org/10.1103/PhysRevLett.119.161101}{\emph{Phys. Rev. Lett.}
  {\bfseries 119} (2017) 161101},
  [\href{https://arxiv.org/abs/1710.05832}{{\ttfamily 1710.05832}}].

\bibitem{Monitor:2017mdv}
{\scshape Virgo, Fermi-GBM, INTEGRAL, LIGO Scientific} collaboration, B.~P.
  Abbott et~al., \emph{{Gravitational Waves and Gamma-rays from a Binary
  Neutron Star Merger: GW170817 and GRB 170817A}},
  \href{https://doi.org/10.3847/2041-8213/aa920c}{\emph{Astrophys. J.}
  {\bfseries 848} (2017) L13},
  [\href{https://arxiv.org/abs/1710.05834}{{\ttfamily 1710.05834}}].

\bibitem{Bettoni:2016mij}
D.~Bettoni, J.~M. Ezquiaga, K.~Hinterbichler and M.~Zumalacárregui,
  \emph{{Speed of Gravitational Waves and the Fate of Scalar-Tensor Gravity}},
  \href{https://doi.org/10.1103/PhysRevD.95.084029}{\emph{Phys. Rev.}
  {\bfseries D95} (2017) 084029},
  [\href{https://arxiv.org/abs/1608.01982}{{\ttfamily 1608.01982}}].

\bibitem{Lombriser:2015sxa}
L.~Lombriser and A.~Taylor, \emph{{Breaking a Dark Degeneracy with
  Gravitational Waves}},
  \href{https://doi.org/10.1088/1475-7516/2016/03/031}{\emph{JCAP} {\bfseries
  1603} (2016) 031}, [\href{https://arxiv.org/abs/1509.08458}{{\ttfamily
  1509.08458}}].

\bibitem{deRham:2016wji}
C.~de~Rham and A.~Matas, \emph{{Ostrogradsky in Theories with Multiple
  Fields}}, \href{https://doi.org/10.1088/1475-7516/2016/06/041}{\emph{JCAP}
  {\bfseries 1606} (2016) 041},
  [\href{https://arxiv.org/abs/1604.08638}{{\ttfamily 1604.08638}}].

\bibitem{Nojiri:2006ri}
S.~Nojiri and S.~D. Odintsov, \emph{{Introduction to modified gravity and
  gravitational alternative for dark energy}},
  \href{https://doi.org/10.1142/S0219887807001928}{\emph{eConf} {\bfseries
  C0602061} (2006) 06}, [\href{https://arxiv.org/abs/hep-th/0601213}{{\ttfamily
  hep-th/0601213}}].

\bibitem{Nojiri:2010wj}
S.~Nojiri and S.~D. Odintsov, \emph{{Unified cosmic history in modified
  gravity: from F(R) theory to Lorentz non-invariant models}},
  \href{https://doi.org/10.1016/j.physrep.2011.04.001}{\emph{Phys. Rept.}
  {\bfseries 505} (2011) 59--144},
  [\href{https://arxiv.org/abs/1011.0544}{{\ttfamily 1011.0544}}].

\bibitem{Clifton:2011jh}
T.~Clifton, P.~G. Ferreira, A.~Padilla and C.~Skordis, \emph{{Modified Gravity
  and Cosmology}},
  \href{https://doi.org/10.1016/j.physrep.2012.01.001}{\emph{Phys. Rept.}
  {\bfseries 513} (2012) 1--189},
  [\href{https://arxiv.org/abs/1106.2476}{{\ttfamily 1106.2476}}].

\bibitem{Capozziello:2011et}
S.~Capozziello and M.~De~Laurentis, \emph{{Extended Theories of Gravity}},
  \href{https://doi.org/10.1016/j.physrep.2011.09.003}{\emph{Phys. Rept.}
  {\bfseries 509} (2011) 167--321},
  [\href{https://arxiv.org/abs/1108.6266}{{\ttfamily 1108.6266}}].

\bibitem{Nojiri:2017ncd}
S.~Nojiri, S.~D. Odintsov and V.~K. Oikonomou, \emph{{Modified Gravity Theories
  on a Nutshell: Inflation, Bounce and Late-time Evolution}},
  \href{https://doi.org/10.1016/j.physrep.2017.06.001}{\emph{Phys. Rept.}
  {\bfseries 692} (2017) 1--104},
  [\href{https://arxiv.org/abs/1705.11098}{{\ttfamily 1705.11098}}].

\bibitem{Creminelli:2017sry}
P.~Creminelli and F.~Vernizzi, \emph{{Dark Energy after GW170817 and
  GRB170817A}},
  \href{https://doi.org/10.1103/PhysRevLett.119.251302}{\emph{Phys. Rev. Lett.}
  {\bfseries 119} (2017) 251302},
  [\href{https://arxiv.org/abs/1710.05877}{{\ttfamily 1710.05877}}].

\bibitem{Sakstein:2017xjx}
J.~Sakstein and B.~Jain, \emph{{Implications of the Neutron Star Merger
  GW170817 for Cosmological Scalar-Tensor Theories}},
  \href{https://doi.org/10.1103/PhysRevLett.119.251303}{\emph{Phys. Rev. Lett.}
  {\bfseries 119} (2017) 251303},
  [\href{https://arxiv.org/abs/1710.05893}{{\ttfamily 1710.05893}}].

\bibitem{Ezquiaga:2017ekz}
J.~M. Ezquiaga and M.~Zumalacárregui, \emph{{Dark Energy After GW170817: Dead
  Ends and the Road Ahead}},
  \href{https://doi.org/10.1103/PhysRevLett.119.251304}{\emph{Phys. Rev. Lett.}
  {\bfseries 119} (2017) 251304},
  [\href{https://arxiv.org/abs/1710.05901}{{\ttfamily 1710.05901}}].

\bibitem{Baker:2017hug}
T.~Baker, E.~Bellini, P.~G. Ferreira, M.~Lagos, J.~Noller and I.~Sawicki,
  \emph{{Strong constraints on cosmological gravity from GW170817 and GRB
  170817A}}, \href{https://doi.org/10.1103/PhysRevLett.119.251301}{\emph{Phys.
  Rev. Lett.} {\bfseries 119} (2017) 251301},
  [\href{https://arxiv.org/abs/1710.06394}{{\ttfamily 1710.06394}}].

\bibitem{Ezquiaga:2018btd}
J.~M. Ezquiaga and M.~Zumalacárregui, \emph{{Dark Energy in light of
  Multi-Messenger Gravitational-Wave astronomy}},
  \href{https://arxiv.org/abs/1807.09241}{{\ttfamily 1807.09241}}.

\bibitem{Lombriser:2016yzn}
L.~Lombriser and N.~A. Lima, \emph{{Challenges to Self-Acceleration in Modified
  Gravity from Gravitational Waves and Large-Scale Structure}},
  \href{https://doi.org/10.1016/j.physletb.2016.12.048}{\emph{Phys. Lett.}
  {\bfseries B765} (2017) 382--385},
  [\href{https://arxiv.org/abs/1602.07670}{{\ttfamily 1602.07670}}].

\bibitem{Boran:2017rdn}
S.~Boran, S.~Desai, E.~O. Kahya and R.~P. Woodard, \emph{{GW170817 Falsifies
  Dark Matter Emulators}},
  \href{https://doi.org/10.1103/PhysRevD.97.041501}{\emph{Phys. Rev.}
  {\bfseries D97} (2018) 041501},
  [\href{https://arxiv.org/abs/1710.06168}{{\ttfamily 1710.06168}}].

\bibitem{Nojiri:2017hai}
S.~Nojiri and S.~D. Odintsov, \emph{{Cosmological Bound from the Neutron Star
  Merger GW170817 in scalar-tensor and $F(R)$ gravity theories}},
  \href{https://doi.org/10.1016/j.physletb.2018.01.078}{\emph{Phys. Lett.}
  {\bfseries B779} (2018) 425--429},
  [\href{https://arxiv.org/abs/1711.00492}{{\ttfamily 1711.00492}}].

\bibitem{Arai:2017hxj}
S.~Arai and A.~Nishizawa, \emph{{Generalized framework for testing gravity with
  gravitational-wave propagation. II. Constraints on Horndeski theory}},
  \href{https://doi.org/10.1103/PhysRevD.97.104038}{\emph{Phys. Rev.}
  {\bfseries D97} (2018) 104038},
  [\href{https://arxiv.org/abs/1711.03776}{{\ttfamily 1711.03776}}].

\bibitem{Amendola:2017orw}
L.~Amendola, M.~Kunz, I.~D. Saltas and I.~Sawicki, \emph{{Fate of Large-Scale
  Structure in Modified Gravity After GW170817 and GRB170817A}},
  \href{https://doi.org/10.1103/PhysRevLett.120.131101}{\emph{Phys. Rev. Lett.}
  {\bfseries 120} (2018) 131101},
  [\href{https://arxiv.org/abs/1711.04825}{{\ttfamily 1711.04825}}].

\bibitem{Visinelli:2017bny}
L.~Visinelli, N.~Bolis and S.~Vagnozzi, \emph{{Brane-world extra dimensions in
  light of GW170817}},
  \href{https://doi.org/10.1103/PhysRevD.97.064039}{\emph{Phys. Rev.}
  {\bfseries D97} (2018) 064039},
  [\href{https://arxiv.org/abs/1711.06628}{{\ttfamily 1711.06628}}].

\bibitem{Crisostomi:2017lbg}
M.~Crisostomi and K.~Koyama, \emph{{Vainshtein mechanism after GW170817}},
  \href{https://doi.org/10.1103/PhysRevD.97.021301}{\emph{Phys. Rev.}
  {\bfseries D97} (2018) 021301},
  [\href{https://arxiv.org/abs/1711.06661}{{\ttfamily 1711.06661}}].

\bibitem{Langlois:2017dyl}
D.~Langlois, R.~Saito, D.~Yamauchi and K.~Noui, \emph{{Scalar-tensor theories
  and modified gravity in the wake of GW170817}},
  \href{https://doi.org/10.1103/PhysRevD.97.061501}{\emph{Phys. Rev.}
  {\bfseries D97} (2018) 061501},
  [\href{https://arxiv.org/abs/1711.07403}{{\ttfamily 1711.07403}}].

\bibitem{Gumrukcuoglu:2017ijh}
A.~Emir~Gümrükçüoğlu, M.~Saravani and T.~P. Sotiriou, \emph{{Hořava
  gravity after GW170817}},
  \href{https://doi.org/10.1103/PhysRevD.97.024032}{\emph{Phys. Rev.}
  {\bfseries D97} (2018) 024032},
  [\href{https://arxiv.org/abs/1711.08845}{{\ttfamily 1711.08845}}].

\bibitem{Kreisch:2017uet}
C.~D. Kreisch and E.~Komatsu, \emph{{Cosmological Constraints on Horndeski
  Gravity in Light of GW170817}},
  \href{https://arxiv.org/abs/1712.02710}{{\ttfamily 1712.02710}}.

\bibitem{Bartolo:2017ibw}
N.~Bartolo, P.~Karmakar, S.~Matarrese and M.~Scomparin, \emph{{Cosmic
  structures and gravitational waves in ghost-free scalar-tensor theories of
  gravity}}, \href{https://doi.org/10.1088/1475-7516/2018/05/048}{\emph{JCAP}
  {\bfseries 1805} (2018) 048},
  [\href{https://arxiv.org/abs/1712.04002}{{\ttfamily 1712.04002}}].

\bibitem{Dima:2017pwp}
A.~Dima and F.~Vernizzi, \emph{{Vainshtein Screening in Scalar-Tensor Theories
  before and after GW170817: Constraints on Theories beyond Horndeski}},
  \href{https://doi.org/10.1103/PhysRevD.97.101302}{\emph{Phys. Rev.}
  {\bfseries D97} (2018) 101302},
  [\href{https://arxiv.org/abs/1712.04731}{{\ttfamily 1712.04731}}].

\bibitem{Cai:2018rzd}
Y.-F. Cai, C.~Li, E.~N. Saridakis and L.~Xue, \emph{{$f(T)$ gravity after
  GW170817 and GRB170817A}},
  \href{https://doi.org/10.1103/PhysRevD.97.103513}{\emph{Phys. Rev.}
  {\bfseries D97} (2018) 103513},
  [\href{https://arxiv.org/abs/1801.05827}{{\ttfamily 1801.05827}}].

\bibitem{Pardo:2018ipy}
K.~Pardo, M.~Fishbach, D.~E. Holz and D.~N. Spergel, \emph{{Limits on the
  number of spacetime dimensions from GW170817}},
  \href{https://doi.org/10.1088/1475-7516/2018/07/048}{\emph{JCAP} {\bfseries
  1807} (2018) 048}, [\href{https://arxiv.org/abs/1801.08160}{{\ttfamily
  1801.08160}}].

\bibitem{Chamseddine:2013kea}
A.~H. Chamseddine and V.~Mukhanov, \emph{{Mimetic Dark Matter}},
  \href{https://doi.org/10.1007/JHEP11(2013)135}{\emph{JHEP} {\bfseries 11}
  (2013) 135}, [\href{https://arxiv.org/abs/1308.5410}{{\ttfamily 1308.5410}}].

\bibitem{Lim:2010yk}
E.~A. Lim, I.~Sawicki and A.~Vikman, \emph{{Dust of Dark Energy}},
  \href{https://doi.org/10.1088/1475-7516/2010/05/012}{\emph{JCAP} {\bfseries
  1005} (2010) 012}, [\href{https://arxiv.org/abs/1003.5751}{{\ttfamily
  1003.5751}}].

\bibitem{Capozziello:2010uv}
S.~Capozziello, J.~Matsumoto, S.~Nojiri and S.~D. Odintsov, \emph{{Dark energy
  from modified gravity with Lagrange multipliers}},
  \href{https://doi.org/10.1016/j.physletb.2010.08.030}{\emph{Phys. Lett.}
  {\bfseries B693} (2010) 198--208},
  [\href{https://arxiv.org/abs/1004.3691}{{\ttfamily 1004.3691}}].

\bibitem{Gao:2010gj}
C.~Gao, Y.~Gong, X.~Wang and X.~Chen, \emph{{Cosmological models with Lagrange
  Multiplier Field}},
  \href{https://doi.org/10.1016/j.physletb.2011.06.085}{\emph{Phys. Lett.}
  {\bfseries B702} (2011) 107--113},
  [\href{https://arxiv.org/abs/1003.6056}{{\ttfamily 1003.6056}}].

\bibitem{Zumalacarregui:2013pma}
M.~Zumalacárregui and J.~García-Bellido, \emph{{Transforming gravity: from
  derivative couplings to matter to second-order scalar-tensor theories beyond
  the Horndeski Lagrangian}},
  \href{https://doi.org/10.1103/PhysRevD.89.064046}{\emph{Phys. Rev.}
  {\bfseries D89} (2014) 064046},
  [\href{https://arxiv.org/abs/1308.4685}{{\ttfamily 1308.4685}}].

\bibitem{Deruelle:2014zza}
N.~Deruelle and J.~Rua, \emph{{Disformal Transformations, Veiled General
  Relativity and Mimetic Gravity}},
  \href{https://doi.org/10.1088/1475-7516/2014/09/002}{\emph{JCAP} {\bfseries
  1409} (2014) 002}, [\href{https://arxiv.org/abs/1407.0825}{{\ttfamily
  1407.0825}}].

\bibitem{Domenech:2015tca}
G.~Domènech, S.~Mukohyama, R.~Namba, A.~Naruko, R.~Saitou and Y.~Watanabe,
  \emph{{Derivative-dependent metric transformation and physical degrees of
  freedom}}, \href{https://doi.org/10.1103/PhysRevD.92.084027}{\emph{Phys.
  Rev.} {\bfseries D92} (2015) 084027},
  [\href{https://arxiv.org/abs/1507.05390}{{\ttfamily 1507.05390}}].

\bibitem{Achour:2016rkg}
J.~Ben~Achour, D.~Langlois and K.~Noui, \emph{{Degenerate higher order
  scalar-tensor theories beyond Horndeski and disformal transformations}},
  \href{https://doi.org/10.1103/PhysRevD.93.124005}{\emph{Phys. Rev.}
  {\bfseries D93} (2016) 124005},
  [\href{https://arxiv.org/abs/1602.08398}{{\ttfamily 1602.08398}}].

\bibitem{Chamseddine:2014vna}
A.~H. Chamseddine, V.~Mukhanov and A.~Vikman, \emph{{Cosmology with Mimetic
  Matter}}, \href{https://doi.org/10.1088/1475-7516/2014/06/017}{\emph{JCAP}
  {\bfseries 1406} (2014) 017},
  [\href{https://arxiv.org/abs/1403.3961}{{\ttfamily 1403.3961}}].

\bibitem{Sebastiani:2016ras}
L.~Sebastiani, S.~Vagnozzi and R.~Myrzakulov, \emph{{Mimetic gravity: a review
  of recent developments and applications to cosmology and astrophysics}},
  \href{https://doi.org/10.1155/2017/3156915}{\emph{Adv. High Energy Phys.}
  {\bfseries 2017} (2017) 3156915},
  [\href{https://arxiv.org/abs/1612.08661}{{\ttfamily 1612.08661}}].

\bibitem{Langlois:2015cwa}
D.~Langlois and K.~Noui, \emph{{Degenerate higher derivative theories beyond
  Horndeski: evading the Ostrogradski instability}},
  \href{https://doi.org/10.1088/1475-7516/2016/02/034}{\emph{JCAP} {\bfseries
  1602} (2016) 034}, [\href{https://arxiv.org/abs/1510.06930}{{\ttfamily
  1510.06930}}].

\bibitem{Langlois:2018jdg}
D.~Langlois, M.~Mancarella, K.~Noui and F.~Vernizzi, \emph{{Mimetic gravity as
  DHOST theories}},  \href{https://arxiv.org/abs/1802.03394}{{\ttfamily
  1802.03394}}.

\bibitem{Cognola:2016gjy}
G.~Cognola, R.~Myrzakulov, L.~Sebastiani, S.~Vagnozzi and S.~Zerbini,
  \emph{{Covariant Ho\v{r}ava-like and mimetic Horndeski gravity: cosmological
  solutions and perturbations}},
  \href{https://doi.org/10.1088/0264-9381/33/22/225014}{\emph{Class. Quant.
  Grav.} {\bfseries 33} (2016) 225014},
  [\href{https://arxiv.org/abs/1601.00102}{{\ttfamily 1601.00102}}].

\bibitem{Golovnev:2013jxa}
A.~Golovnev, \emph{{On the recently proposed Mimetic Dark Matter}},
  \href{https://doi.org/10.1016/j.physletb.2013.11.026}{\emph{Phys. Lett.}
  {\bfseries B728} (2014) 39--40},
  [\href{https://arxiv.org/abs/1310.2790}{{\ttfamily 1310.2790}}].

\bibitem{Barvinsky:2013mea}
A.~O. Barvinsky, \emph{{Dark matter as a ghost free conformal extension of
  Einstein theory}},
  \href{https://doi.org/10.1088/1475-7516/2014/01/014}{\emph{JCAP} {\bfseries
  1401} (2014) 014}, [\href{https://arxiv.org/abs/1311.3111}{{\ttfamily
  1311.3111}}].

\bibitem{Nojiri:2014zqa}
S.~Nojiri and S.~D. Odintsov, \emph{{Mimetic $F(R)$ gravity: inflation, dark
  energy and bounce}},
  \href{https://doi.org/10.1142/S0217732314502113}{\emph{Mod. Phys. Lett.}
  {\bfseries A29} (2014) 1450211},
  [\href{https://arxiv.org/abs/1408.3561}{{\ttfamily 1408.3561}}].

\bibitem{Saadi:2014jfa}
H.~Saadi, \emph{{A Cosmological Solution to Mimetic Dark Matter}},
  \href{https://doi.org/10.1140/epjc/s10052-015-3856-0}{\emph{Eur. Phys. J.}
  {\bfseries C76} (2016) 14},
  [\href{https://arxiv.org/abs/1411.4531}{{\ttfamily 1411.4531}}].

\bibitem{Capela:2014xta}
F.~Capela and S.~Ramazanov, \emph{{Modified Dust and the Small Scale Crisis in
  CDM}}, \href{https://doi.org/10.1088/1475-7516/2015/04/051}{\emph{JCAP}
  {\bfseries 1504} (2015) 051},
  [\href{https://arxiv.org/abs/1412.2051}{{\ttfamily 1412.2051}}].

\bibitem{Mirzagholi:2014ifa}
L.~Mirzagholi and A.~Vikman, \emph{{Imperfect Dark Matter}},
  \href{https://doi.org/10.1088/1475-7516/2015/06/028}{\emph{JCAP} {\bfseries
  1506} (2015) 028}, [\href{https://arxiv.org/abs/1412.7136}{{\ttfamily
  1412.7136}}].

\bibitem{Leon:2014yua}
G.~Leon and E.~N. Saridakis, \emph{{Dynamical behavior in mimetic F(R)
  gravity}}, \href{https://doi.org/10.1088/1475-7516/2015/04/031}{\emph{JCAP}
  {\bfseries 1504} (2015) 031},
  [\href{https://arxiv.org/abs/1501.00488}{{\ttfamily 1501.00488}}].

\bibitem{Haghani:2015iva}
Z.~Haghani, T.~Harko, H.~R. Sepangi and S.~Shahidi, \emph{{Cosmology of a
  Lorentz violating Galileon theory}},
  \href{https://doi.org/10.1088/1475-7516/2015/05/022}{\emph{JCAP} {\bfseries
  1505} (2015) 022}, [\href{https://arxiv.org/abs/1501.00819}{{\ttfamily
  1501.00819}}].

\bibitem{Matsumoto:2015wja}
J.~Matsumoto, S.~D. Odintsov and S.~V. Sushkov, \emph{{Cosmological
  perturbations in a mimetic matter model}},
  \href{https://doi.org/10.1103/PhysRevD.91.064062}{\emph{Phys. Rev.}
  {\bfseries D91} (2015) 064062},
  [\href{https://arxiv.org/abs/1501.02149}{{\ttfamily 1501.02149}}].

\bibitem{Momeni:2015gka}
D.~Momeni, R.~Myrzakulov and E.~Güdekli, \emph{{Cosmological viable mimetic
  $f(R)$ and $f(R,T)$ theories via Noether symmetry}},
  \href{https://doi.org/10.1142/S0219887815501017}{\emph{Int. J. Geom. Meth.
  Mod. Phys.} {\bfseries 12} (2015) 1550101},
  [\href{https://arxiv.org/abs/1502.00977}{{\ttfamily 1502.00977}}].

\bibitem{Myrzakulov:2015sea}
R.~Myrzakulov and L.~Sebastiani, \emph{{Spherically symmetric static vacuum
  solutions in Mimetic gravity}},
  \href{https://doi.org/10.1007/s10714-015-1930-4}{\emph{Gen. Rel. Grav.}
  {\bfseries 47} (2015) 89},
  [\href{https://arxiv.org/abs/1503.04293}{{\ttfamily 1503.04293}}].

\bibitem{Astashenok:2015haa}
A.~V. Astashenok, S.~D. Odintsov and V.~K. Oikonomou, \emph{{Modified
  Gauss–Bonnet gravity with the Lagrange multiplier constraint as mimetic
  theory}}, \href{https://doi.org/10.1088/0264-9381/32/18/185007}{\emph{Class.
  Quant. Grav.} {\bfseries 32} (2015) 185007},
  [\href{https://arxiv.org/abs/1504.04861}{{\ttfamily 1504.04861}}].

\bibitem{Myrzakulov:2015qaa}
R.~Myrzakulov, L.~Sebastiani and S.~Vagnozzi, \emph{{Inflation in $f(R,\phi )$
  -theories and mimetic gravity scenario}},
  \href{https://doi.org/10.1140/epjc/s10052-015-3672-6}{\emph{Eur. Phys. J.}
  {\bfseries C75} (2015) 444},
  [\href{https://arxiv.org/abs/1504.07984}{{\ttfamily 1504.07984}}].

\bibitem{Myrzakulov:2015kda}
R.~Myrzakulov, L.~Sebastiani, S.~Vagnozzi and S.~Zerbini, \emph{{Static
  spherically symmetric solutions in mimetic gravity: rotation curves and
  wormholes}},
  \href{https://doi.org/10.1088/0264-9381/33/12/125005}{\emph{Class. Quant.
  Grav.} {\bfseries 33} (2016) 125005},
  [\href{https://arxiv.org/abs/1510.02284}{{\ttfamily 1510.02284}}].

\bibitem{Odintsov:2015wwp}
S.~D. Odintsov and V.~K. Oikonomou, \emph{{Accelerating cosmologies and the
  phase structure of F(R) gravity with Lagrange multiplier constraints: A
  mimetic approach}},
  \href{https://doi.org/10.1103/PhysRevD.93.023517}{\emph{Phys. Rev.}
  {\bfseries D93} (2016) 023517},
  [\href{https://arxiv.org/abs/1511.04559}{{\ttfamily 1511.04559}}].

\bibitem{Hammer:2015pcx}
K.~Hammer and A.~Vikman, \emph{{Many Faces of Mimetic Gravity}},
  \href{https://arxiv.org/abs/1512.09118}{{\ttfamily 1512.09118}}.

\bibitem{Ramazanov:2016xhp}
S.~Ramazanov, F.~Arroja, M.~Celoria, S.~Matarrese and L.~Pilo, \emph{{Living
  with ghosts in Hořava-Lifshitz gravity}},
  \href{https://doi.org/10.1007/JHEP06(2016)020}{\emph{JHEP} {\bfseries 06}
  (2016) 020}, [\href{https://arxiv.org/abs/1601.05405}{{\ttfamily
  1601.05405}}].

\bibitem{Nojiri:2016ppu}
S.~Nojiri, S.~D. Odintsov and V.~K. Oikonomou, \emph{{Unimodular-Mimetic
  Cosmology}},
  \href{https://doi.org/10.1088/0264-9381/33/12/125017}{\emph{Class. Quant.
  Grav.} {\bfseries 33} (2016) 125017},
  [\href{https://arxiv.org/abs/1601.07057}{{\ttfamily 1601.07057}}].

\bibitem{Nojiri:2016vhu}
S.~Nojiri, S.~Odintsov and V.~Oikonomou, \emph{{Viable Mimetic Completion of
  Unified Inflation-Dark Energy Evolution in Modified Gravity}},
  \href{https://doi.org/10.1103/PhysRevD.94.104050}{\emph{Phys. Rev.}
  {\bfseries D94} (2016) 104050},
  [\href{https://arxiv.org/abs/1608.07806}{{\ttfamily 1608.07806}}].

\bibitem{Sadeghnezhad:2017hmr}
N.~Sadeghnezhad and K.~Nozari, \emph{{Braneworld Mimetic Cosmology}},
  \href{https://doi.org/10.1016/j.physletb.2017.03.039}{\emph{Phys. Lett.}
  {\bfseries B769} (2017) 134--140},
  [\href{https://arxiv.org/abs/1703.06269}{{\ttfamily 1703.06269}}].

\bibitem{Baffou:2017pao}
E.~H. Baffou, M.~J.~S. Houndjo, M.~Hamani-Daouda and F.~G. Alvarenga,
  \emph{{Late time cosmological approach in mimetic $f(R,T)$ gravity}},
  \href{https://doi.org/10.1140/epjc/s10052-017-5291-x}{\emph{Eur. Phys. J.}
  {\bfseries C77} (2017) 708},
  [\href{https://arxiv.org/abs/1706.08842}{{\ttfamily 1706.08842}}].

\bibitem{Vagnozzi:2017ilo}
S.~Vagnozzi, \emph{{Recovering a MOND-like acceleration law in mimetic
  gravity}}, \href{https://doi.org/10.1088/1361-6382/aa838b}{\emph{Class.
  Quant. Grav.} {\bfseries 34} (2017) 185006},
  [\href{https://arxiv.org/abs/1708.00603}{{\ttfamily 1708.00603}}].

\bibitem{Bouhmadi-Lopez:2017lbx}
M.~Bouhmadi-López, C.-Y. Chen and P.~Chen, \emph{{Primordial Cosmology in
  Mimetic Born-Infeld Gravity}},
  \href{https://doi.org/10.1088/1475-7516/2017/11/053}{\emph{JCAP} {\bfseries
  1711} (2017) 053}, [\href{https://arxiv.org/abs/1709.09192}{{\ttfamily
  1709.09192}}].

\bibitem{Shen:2017rya}
L.~Shen, Y.~Mou, Y.~Zheng and M.~Li, \emph{{Direct couplings of mimetic dark
  matter and their cosmological effects}},
  \href{https://doi.org/10.1088/1674-1137/42/1/015101}{\emph{Chin. Phys.}
  {\bfseries C42} (2018) 015101},
  [\href{https://arxiv.org/abs/1710.03945}{{\ttfamily 1710.03945}}].

\bibitem{Nojiri:2017ygt}
S.~Nojiri, S.~D. Odintsov and V.~K. Oikonomou, \emph{{Ghost-Free $F(R)$ Gravity
  with Lagrange Multiplier Constraint}},
  \href{https://doi.org/10.1016/j.physletb.2017.10.045}{\emph{Phys. Lett.}
  {\bfseries B775} (2017) 44--49},
  [\href{https://arxiv.org/abs/1710.07838}{{\ttfamily 1710.07838}}].

\bibitem{Dutta:2017fjw}
J.~Dutta, W.~Khyllep, E.~N. Saridakis, N.~Tamanini and S.~Vagnozzi,
  \emph{{Cosmological dynamics of mimetic gravity}},
  \href{https://doi.org/10.1088/1475-7516/2018/02/041}{\emph{JCAP} {\bfseries
  1802} (2018) 041}, [\href{https://arxiv.org/abs/1711.07290}{{\ttfamily
  1711.07290}}].

\bibitem{Brahma:2018dwx}
S.~Brahma, A.~Golovnev and D.-H. Yeom, \emph{{On singularity-resolution in
  mimetic gravity}},
  \href{https://doi.org/10.1016/j.physletb.2018.05.039}{\emph{Phys. Lett.}
  {\bfseries B782} (2018) 280--284},
  [\href{https://arxiv.org/abs/1803.03955}{{\ttfamily 1803.03955}}].

\bibitem{deHaro:2018sqw}
J.~de~Haro, L.~Aresté~Saló and S.~Pan, \emph{{Mimetic Loop Quantum
  Cosmology}},  \href{https://arxiv.org/abs/1803.09653}{{\ttfamily
  1803.09653}}.

\bibitem{Zhong:2018tqn}
Y.~Zhong and D.~Sáez-Chillón~Gómez, \emph{{Inflation in mimetic $f(G)$
  gravity}}, \href{https://doi.org/10.3390/sym10050170}{\emph{Symmetry}
  {\bfseries 10} (2018) 170},
  [\href{https://arxiv.org/abs/1805.03467}{{\ttfamily 1805.03467}}].

\bibitem{Chamseddine:2018qym}
A.~H. Chamseddine and V.~Mukhanov, \emph{{Ghost Free Mimetic Massive Gravity}},
  \href{https://doi.org/10.1007/JHEP06(2018)060}{\emph{JHEP} {\bfseries 06}
  (2018) 060}, [\href{https://arxiv.org/abs/1805.06283}{{\ttfamily
  1805.06283}}].

\bibitem{Chamseddine:2018gqh}
A.~H. Chamseddine and V.~Mukhanov, \emph{{Mimetic Massive Gravity: Beyond
  Linear Approximation}},
  \href{https://doi.org/10.1007/JHEP06(2018)062}{\emph{JHEP} {\bfseries 06}
  (2018) 062}, [\href{https://arxiv.org/abs/1805.06598}{{\ttfamily
  1805.06598}}].

\bibitem{Firouzjahi:2018xob}
H.~Firouzjahi, M.~A. Gorji, S.~A. Hosseini~Mansoori, A.~Karami and T.~Rostami,
  \emph{{Two-field disformal transformation and mimetic cosmology}},
  \href{https://arxiv.org/abs/1806.11472}{{\ttfamily 1806.11472}}.

\bibitem{Zlosnik:2018qvg}
T.~Złośnik, F.~Urban, L.~Marzola and T.~Koivisto, \emph{{Spacetime and dark
  matter from spontaneous breaking of Lorentz symmetry}},
  \href{https://arxiv.org/abs/1807.01100}{{\ttfamily 1807.01100}}.

\bibitem{Gorji:2018okn}
M.~A. Gorji, S.~Mukohyama, H.~Firouzjahi and S.~A. Hosseini~Mansoori,
  \emph{{Gauge Field Mimetic Cosmology}},
  \href{https://doi.org/10.1088/1475-7516/2018/08/047}{\emph{JCAP} {\bfseries
  1808} (2018) 047}, [\href{https://arxiv.org/abs/1807.06335}{{\ttfamily
  1807.06335}}].

\bibitem{Nashed:2018qag}
G.~G.~L. Nashed, W.~El~Hanafy and K.~Bamba, \emph{{Charged rotating black holes
  coupled with nonlinear electrodynamics Maxwell field in the mimetic
  gravity}},  \href{https://arxiv.org/abs/1809.02289}{{\ttfamily 1809.02289}}.

\bibitem{Ganz:2018vzg}
A.~Ganz, N.~Bartolo, P.~Karmakar and S.~Matarrese, \emph{{Gravity in mimetic
  scalar-tensor theories after GW170817}},
  \href{https://arxiv.org/abs/1809.03496}{{\ttfamily 1809.03496}}.

\bibitem{Horndeski:1974wa}
G.~W. Horndeski, \emph{{Second-order scalar-tensor field equations in a
  four-dimensional space}},
  \href{https://doi.org/10.1007/BF01807638}{\emph{Int. J. Theor. Phys.}
  {\bfseries 10} (1974) 363--384}.

\bibitem{Arroja:2015wpa}
F.~Arroja, N.~Bartolo, P.~Karmakar and S.~Matarrese, \emph{{The two faces of
  mimetic Horndeski gravity: disformal transformations and Lagrange
  multiplier}},
  \href{https://doi.org/10.1088/1475-7516/2015/09/051}{\emph{JCAP} {\bfseries
  1509} (2015) 051}, [\href{https://arxiv.org/abs/1506.08575}{{\ttfamily
  1506.08575}}].

\bibitem{Rabochaya:2015haa}
Y.~Rabochaya and S.~Zerbini, \emph{{A note on a mimetic scalar–tensor
  cosmological model}},
  \href{https://doi.org/10.1140/epjc/s10052-016-3926-y}{\emph{Eur. Phys. J.}
  {\bfseries C76} (2016) 85},
  [\href{https://arxiv.org/abs/1509.03720}{{\ttfamily 1509.03720}}].

\bibitem{Arroja:2015yvd}
F.~Arroja, N.~Bartolo, P.~Karmakar and S.~Matarrese, \emph{{Cosmological
  perturbations in mimetic Horndeski gravity}},
  \href{https://doi.org/10.1088/1475-7516/2016/04/042}{\emph{JCAP} {\bfseries
  1604} (2016) 042}, [\href{https://arxiv.org/abs/1512.09374}{{\ttfamily
  1512.09374}}].

\bibitem{Arroja:2017msd}
F.~Arroja, T.~Okumura, N.~Bartolo, P.~Karmakar and S.~Matarrese,
  \emph{{Large-scale structure in mimetic Horndeski gravity}},
  \href{https://arxiv.org/abs/1708.01850}{{\ttfamily 1708.01850}}.

\bibitem{jirousekthesis}
P.~Jirou\v{s}ek, \emph{{On Extended Mimetic Gravity}},  Master's thesis,
  Charles U., 2016-09-08.

\bibitem{Rinaldi:2016oqp}
M.~Rinaldi, \emph{{Mimicking dark matter in Horndeski gravity}},
  \href{https://doi.org/10.1016/j.dark.2017.02.003}{\emph{Phys. Dark Univ.}
  {\bfseries 16} (2017) 14--21},
  [\href{https://arxiv.org/abs/1608.03839}{{\ttfamily 1608.03839}}].

\bibitem{Diez-Tejedor:2018fue}
A.~Diez-Tejedor, F.~Flores and G.~Niz, \emph{{Horndeski dark matter and
  beyond}}, \href{https://doi.org/10.1103/PhysRevD.97.123524}{\emph{Phys. Rev.}
  {\bfseries D97} (2018) 123524},
  [\href{https://arxiv.org/abs/1803.00014}{{\ttfamily 1803.00014}}].

\bibitem{Casalino:2018tcd}
A.~Casalino, M.~Rinaldi, L.~Sebastiani and S.~Vagnozzi, \emph{{Mimicking dark
  matter and dark energy in a mimetic model compatible with GW170817}},
  \href{https://doi.org/10.1016/j.dark.2018.10.001}{\emph{Phys. Dark Univ.}
  {\bfseries 22} (2018) 108--115},
  [\href{https://arxiv.org/abs/1803.02620}{{\ttfamily 1803.02620}}].

\bibitem{Horava:2009uw}
P.~Horava, \emph{{Quantum Gravity at a Lifshitz Point}},
  \href{https://doi.org/10.1103/PhysRevD.79.084008}{\emph{Phys. Rev.}
  {\bfseries D79} (2009) 084008},
  [\href{https://arxiv.org/abs/0901.3775}{{\ttfamily 0901.3775}}].

\bibitem{Nojiri:2009th}
S.~Nojiri and S.~D. Odintsov, \emph{{Covariant Horava-like renormalizable
  gravity and its FRW cosmology}},
  \href{https://doi.org/10.1103/PhysRevD.81.043001}{\emph{Phys. Rev.}
  {\bfseries D81} (2010) 043001},
  [\href{https://arxiv.org/abs/0905.4213}{{\ttfamily 0905.4213}}].

\bibitem{Nojiri:2010tv}
S.~Nojiri and S.~D. Odintsov, \emph{{A proposal for covariant renormalizable
  field theory of gravity}},
  \href{https://doi.org/10.1016/j.physletb.2010.06.007}{\emph{Phys. Lett.}
  {\bfseries B691} (2010) 60--64},
  [\href{https://arxiv.org/abs/1004.3613}{{\ttfamily 1004.3613}}].

\bibitem{Cognola:2010by}
G.~Cognola, E.~Elizalde, L.~Sebastiani and S.~Zerbini, \emph{{Black hole and de
  Sitter solutions in a covariant renormalizable field theory of gravity}},
  \href{https://doi.org/10.1103/PhysRevD.83.063003}{\emph{Phys. Rev.}
  {\bfseries D83} (2011) 063003},
  [\href{https://arxiv.org/abs/1007.4676}{{\ttfamily 1007.4676}}].

\bibitem{Nojiri:2010kx}
S.~Nojiri and S.~D. Odintsov, \emph{{Covariant power-counting renormalizable
  gravity: Lorentz symmetry breaking and accelerating early-time FRW
  universe}}, \href{https://doi.org/10.1103/PhysRevD.83.023001}{\emph{Phys.
  Rev.} {\bfseries D83} (2011) 023001},
  [\href{https://arxiv.org/abs/1007.4856}{{\ttfamily 1007.4856}}].

\bibitem{Chaichian:2011sx}
M.~Chaichian, M.~Oksanen and A.~Tureanu, \emph{{Arnowitt-Deser-Misner
  representation and Hamiltonian analysis of covariant renormalizable
  gravity}}, \href{https://doi.org/10.1140/epjc/s10052-011-1657-7,
  10.1140/epjc/s10052-011-1736-9}{\emph{Eur. Phys. J.} {\bfseries C71} (2011)
  1657}, [\href{https://arxiv.org/abs/1101.2843}{{\ttfamily 1101.2843}}].

\bibitem{Blas:2009yd}
D.~Blas, O.~Pujolas and S.~Sibiryakov, \emph{{On the Extra Mode and
  Inconsistency of Horava Gravity}},
  \href{https://doi.org/10.1088/1126-6708/2009/10/029}{\emph{JHEP} {\bfseries
  10} (2009) 029}, [\href{https://arxiv.org/abs/0906.3046}{{\ttfamily
  0906.3046}}].

\bibitem{Blas:2009qj}
D.~Blas, O.~Pujolas and S.~Sibiryakov, \emph{{Consistent Extension of Horava
  Gravity}}, \href{https://doi.org/10.1103/PhysRevLett.104.181302}{\emph{Phys.
  Rev. Lett.} {\bfseries 104} (2010) 181302},
  [\href{https://arxiv.org/abs/0909.3525}{{\ttfamily 0909.3525}}].

\bibitem{Blas:2009ck}
D.~Blas, O.~Pujolas and S.~Sibiryakov, \emph{{Comment on `Strong coupling in
  extended Horava-Lifshitz gravity'}},
  \href{https://doi.org/10.1016/j.physletb.2010.03.073}{\emph{Phys. Lett.}
  {\bfseries B688} (2010) 350--355},
  [\href{https://arxiv.org/abs/0912.0550}{{\ttfamily 0912.0550}}].

\bibitem{Adams:2006sv}
A.~Adams, N.~Arkani-Hamed, S.~Dubovsky, A.~Nicolis and R.~Rattazzi,
  \emph{{Causality, analyticity and an IR obstruction to UV completion}},
  \href{https://doi.org/10.1088/1126-6708/2006/10/014}{\emph{JHEP} {\bfseries
  10} (2006) 014}, [\href{https://arxiv.org/abs/hep-th/0602178}{{\ttfamily
  hep-th/0602178}}].

\bibitem{Salvatelli:2016mgy}
V.~Salvatelli, F.~Piazza and C.~Marinoni, \emph{{Constraints on modified
  gravity from Planck 2015: when the health of your theory makes the
  difference}},
  \href{https://doi.org/10.1088/1475-7516/2016/09/027}{\emph{JCAP} {\bfseries
  1609} (2016) 027}, [\href{https://arxiv.org/abs/1602.08283}{{\ttfamily
  1602.08283}}].

\bibitem{Moore:2001bv}
G.~D. Moore and A.~E. Nelson, \emph{{Lower bound on the propagation speed of
  gravity from gravitational Cherenkov radiation}},
  \href{https://doi.org/10.1088/1126-6708/2001/09/023}{\emph{JHEP} {\bfseries
  09} (2001) 023}, [\href{https://arxiv.org/abs/hep-ph/0106220}{{\ttfamily
  hep-ph/0106220}}].

\bibitem{Kunz:2016yqy}
M.~Kunz, S.~Nesseris and I.~Sawicki, \emph{{Constraints on dark-matter
  properties from large-scale structure}},
  \href{https://doi.org/10.1103/PhysRevD.94.023510}{\emph{Phys. Rev.}
  {\bfseries D94} (2016) 023510},
  [\href{https://arxiv.org/abs/1604.05701}{{\ttfamily 1604.05701}}].

\bibitem{Audren:2012wb}
B.~Audren, J.~Lesgourgues, K.~Benabed and S.~Prunet, \emph{{Conservative
  Constraints on Early Cosmology: an illustration of the Monte Python
  cosmological parameter inference code}},
  \href{https://doi.org/10.1088/1475-7516/2013/02/001}{\emph{JCAP} {\bfseries
  1302} (2013) 001}, [\href{https://arxiv.org/abs/1210.7183}{{\ttfamily
  1210.7183}}].

\bibitem{Gelman:1992zz}
A.~Gelman and D.~B. Rubin, \emph{{Inference from Iterative Simulation Using
  Multiple Sequences}},
  \href{https://doi.org/10.1214/ss/1177011136}{\emph{Statist. Sci.} {\bfseries
  7} (1992) 457--472}.

\bibitem{Brouzakis:2013lla}
N.~Brouzakis, A.~Codello, N.~Tetradis and O.~Zanusso, \emph{{Quantum
  corrections in Galileon theories}},
  \href{https://doi.org/10.1103/PhysRevD.89.125017}{\emph{Phys. Rev.}
  {\bfseries D89} (2014) 125017},
  [\href{https://arxiv.org/abs/1310.0187}{{\ttfamily 1310.0187}}].

\bibitem{Saltas:2016nkg}
I.~D. Saltas and V.~Vitagliano, \emph{{Covariantly Quantum Galileon}},
  \href{https://doi.org/10.1103/PhysRevD.95.105002}{\emph{Phys. Rev.}
  {\bfseries D95} (2017) 105002},
  [\href{https://arxiv.org/abs/1611.07984}{{\ttfamily 1611.07984}}].

\bibitem{Saltas:2016awg}
I.~D. Saltas and V.~Vitagliano, \emph{{Quantum corrections for the cubic
  Galileon in the covariant language}},
  \href{https://doi.org/10.1088/1475-7516/2017/05/020}{\emph{JCAP} {\bfseries
  1705} (2017) 020}, [\href{https://arxiv.org/abs/1612.08953}{{\ttfamily
  1612.08953}}].

\bibitem{Chaichian:2014qba}
M.~Chaichian, J.~Kluson, M.~Oksanen and A.~Tureanu, \emph{{Mimetic dark matter,
  ghost instability and a mimetic tensor-vector-scalar gravity}},
  \href{https://doi.org/10.1007/JHEP12(2014)102}{\emph{JHEP} {\bfseries 12}
  (2014) 102}, [\href{https://arxiv.org/abs/1404.4008}{{\ttfamily 1404.4008}}].

\bibitem{Ijjas:2016pad}
A.~Ijjas, J.~Ripley and P.~J. Steinhardt, \emph{{NEC violation in mimetic
  cosmology revisited}},
  \href{https://doi.org/10.1016/j.physletb.2016.06.052}{\emph{Phys. Lett.}
  {\bfseries B760} (2016) 132--138},
  [\href{https://arxiv.org/abs/1604.08586}{{\ttfamily 1604.08586}}].

\bibitem{Firouzjahi:2017txv}
H.~Firouzjahi, M.~A. Gorji and S.~A. Hosseini~Mansoori, \emph{{Instabilities in
  Mimetic Matter Perturbations}},
  \href{https://doi.org/10.1088/1475-7516/2017/07/031}{\emph{JCAP} {\bfseries
  1707} (2017) 031}, [\href{https://arxiv.org/abs/1703.02923}{{\ttfamily
  1703.02923}}].

\bibitem{Yoshida:2017swb}
D.~Yoshida, J.~Quintin, M.~Yamaguchi and R.~H. Brandenberger,
  \emph{{Cosmological perturbations and stability of nonsingular cosmologies
  with limiting curvature}},
  \href{https://doi.org/10.1103/PhysRevD.96.043502}{\emph{Phys. Rev.}
  {\bfseries D96} (2017) 043502},
  [\href{https://arxiv.org/abs/1704.04184}{{\ttfamily 1704.04184}}].

\bibitem{Hirano:2017zox}
S.~Hirano, S.~Nishi and T.~Kobayashi, \emph{{Healthy imperfect dark matter from
  effective theory of mimetic cosmological perturbations}},
  \href{https://doi.org/10.1088/1475-7516/2017/07/009}{\emph{JCAP} {\bfseries
  1707} (2017) 009}, [\href{https://arxiv.org/abs/1704.06031}{{\ttfamily
  1704.06031}}].

\bibitem{Zheng:2017qfs}
Y.~Zheng, L.~Shen, Y.~Mou and M.~Li, \emph{{On (in)stabilities of perturbations
  in mimetic models with higher derivatives}},
  \href{https://doi.org/10.1088/1475-7516/2017/08/040}{\emph{JCAP} {\bfseries
  1708} (2017) 040}, [\href{https://arxiv.org/abs/1704.06834}{{\ttfamily
  1704.06834}}].

\bibitem{Cai:2017dyi}
Y.~Cai and Y.-S. Piao, \emph{{A covariant Lagrangian for stable nonsingular
  bounce}}, \href{https://doi.org/10.1007/JHEP09(2017)027}{\emph{JHEP}
  {\bfseries 09} (2017) 027},
  [\href{https://arxiv.org/abs/1705.03401}{{\ttfamily 1705.03401}}].

\bibitem{Cai:2017dxl}
Y.~Cai and Y.-S. Piao, \emph{{Higher order derivative coupling to gravity and
  its cosmological implications}},
  \href{https://doi.org/10.1103/PhysRevD.96.124028}{\emph{Phys. Rev.}
  {\bfseries D96} (2017) 124028},
  [\href{https://arxiv.org/abs/1707.01017}{{\ttfamily 1707.01017}}].

\bibitem{Takahashi:2017pje}
K.~Takahashi and T.~Kobayashi, \emph{{Extended mimetic gravity: Hamiltonian
  analysis and gradient instabilities}},
  \href{https://doi.org/10.1088/1475-7516/2017/11/038}{\emph{JCAP} {\bfseries
  1711} (2017) 038}, [\href{https://arxiv.org/abs/1708.02951}{{\ttfamily
  1708.02951}}].

\bibitem{Gorji:2017cai}
M.~A. Gorji, S.~A. Hosseini~Mansoori and H.~Firouzjahi, \emph{{Higher
  Derivative Mimetic Gravity}},
  \href{https://doi.org/10.1088/1475-7516/2018/01/020}{\emph{JCAP} {\bfseries
  1801} (2018) 020}, [\href{https://arxiv.org/abs/1709.09988}{{\ttfamily
  1709.09988}}].

\bibitem{Nunes:2018evm}
R.~C. Nunes, S.~Pan and E.~N. Saridakis, \emph{{New observational constraints
  on $f(T)$ gravity through gravitational-wave astronomy}},
  \href{https://arxiv.org/abs/1810.03942}{{\ttfamily 1810.03942}}.

\end{thebibliography}\endgroup

\end{document}